\documentclass[aps,twocolumn,showpacs,superscriptaddress,prm]{revtex4-1}

\usepackage{graphicx}
\usepackage{wrapfig}
\usepackage{amsmath}
\usepackage[per-mode=symbol]{siunitx}

\usepackage{color}
\usepackage[colorlinks,urlcolor=blue,citecolor=blue]{hyperref}
\usepackage[dvipsnames]{xcolor}
\usepackage{tikz}

\begin{document}

\title{First-principles simulation of light-ion microscopy of graphene}

\author{Alina Kononov}
\affiliation{Center for Computing Research, Sandia National Laboratories, Albuquerque NM, USA}

\author{Alexandra Olmstead}
\affiliation{Center for Computing Research, Sandia National Laboratories, Albuquerque NM, USA}

\author{Andrew D. Baczewski}
\affiliation{Center for Computing Research, Sandia National Laboratories, Albuquerque NM, USA}

\author{Andr\'e Schleife}
\affiliation{Department of Materials Science and
Engineering, University of Illinois at Urbana-Champaign, Urbana IL, USA}
\affiliation{Materials Research Laboratory, University of Illinois at Urbana-Champaign, Urbana, IL, USA}
\affiliation{National Center for Supercomputing Applications, University of Illinois at Urbana-Champaign, Urbana, IL, USA}

\begin{abstract}
The extreme sensitivity of 2D materials to defects and nanostructure requires precise imaging techniques to verify presence of desirable and absence of undesirable features in the atomic geometry.
Helium-ion beams have emerged as a promising materials imaging tool, achieving up to 20 times higher resolution and 10 times larger depth-of-field than conventional or environmental scanning electron microscopes.
Here, we offer first-principles theoretical insights to advance ion-beam imaging of atomically thin materials by performing real-time time-dependent density functional theory simulations of single impacts of 10\,--\,\SI{200}{\kilo\electronvolt} light ions in free-standing graphene.
We predict that detecting electrons emitted from the back of the material (the side from which the ion exits) would result in up to 3 times higher signal and up to 5 times higher contrast images, making 2D materials especially compelling targets for ion-beam microscopy.
We also find that the charge induced in the graphene equilibrates on a sub-fs time scale, leading to only slight disturbances in the carbon lattice that are unlikely to damage the atomic structure for any of the beam parameters investigated here.
\end{abstract}

\maketitle

\section{Introduction}
\label{sec:intro}

Recent advances in materials imaging techniques achieve sub-nm resolution by exploiting the shorter deBroglie wavelength and narrower interaction volumes of light ions compared to more common electron and optical microscopy methods \cite{ward:2006,notte:2007,hlawacek:2014,fox:2013,wirtz:2019}.
High-resolution imaging is especially important for 2D materials, in which unwanted defects destroy intrinsic properties \cite{chen:2009,tsen:2012,vicarelli:2015} but intentional structural features including point defects, functional groups, nanopores, and extended defects enable diverse applications \cite{li:2017,vogl:2018,surwade:2015}.
Precise, nondestructive characterization techniques capable of atomic resolution are thus critical for scalable fabrication of devices based on 2D materials.

Depending on ion species, charge, energy, and fluence/dose, focused ion beams can also damage or modify the atomic structure of a material.
In graphene alone, experiments have demonstrated a wide range of ion-induced structural changes, including doping or ion implantation \cite{zhao:2012,bangert:2013,nanda:2015}, cutting or patterning \cite{bell:2009,archanjo:2014,naitou:2015,iberi:2015}, amorphization \cite{pan:2014, kotakoski:2015}, and formation of point defects like reconstructed vacancies \cite{ugeda:2012,lehtinen:2014} and Stone-Wales defects \cite{pan:2014}.
Some of these examples \cite{nanda:2015,bell:2009,archanjo:2014,pan:2014,naitou:2015,iberi:2015} even used the same type of light-ion irradiation, \SI{30}{\kilo\electronvolt} He$^{+}$, as typically employed in microscopy.
Molecular dynamics simulations of ion-irradiated 2D materials \cite{lehtinen:2010,lehtinen:2011:gr,kretschmer:2018,vazquez:2017,lehtinen:2011:bn,ghorbani:2017,ghaderzadeh:2021,kretschmer:2022} have offered some insight into the underlying damage mechanisms, particularly for very slow ions, but these treatments usually neglect electronic excitations which accompany charge transfer processes and dominate energy transfer in the keV\,--\,MeV ion energy regime \cite{ziegler:2010}.

Computational modeling of electron dynamics during ion irradiation of materials offers opportunities to gain insight into the role of excited electrons in damage processes and accurately predict optimal beam and detector parameters for nondestructive imaging of 2D materials.
Many studies have demonstrated the ability of first-principles calculations to predict accurate energy deposition rates for ions traversing bulk materials \cite{pruneda:2007,schleife:2015,yost:2016,lee:2018,lim:2016,quashie:2016,halliday:2019}.
However, the projectile charge may not fully equilibrate within a thin target \cite{wilhelm:2014,wilhelm:2017,wilhelm:2018}, fundamentally altering the response of these materials to ion irradiation.
More recently, several first-principles studies considered ion-irradiated surfaces and 2D materials \cite{kononov:2020,zhang:2012,ojanpera:2014,zhao:2015,kononov:2021,vazquez:2021}, in some cases predicting enhanced energy deposition compared to bulk caused by surface plasmon excitations \cite{kononov:2020} or mediated by projectile charge capture processes \cite{kononov:2021}.
Since energy deposition rates influence an ion beam's ability to damage a sample, damage processes may differ considerably between 2D and bulk materials, requiring special efforts to adapt imaging techniques for the former.

However, even if the energy deposited in the electronic system of a material exceeds defect formation energies, as can occur for single proton and He ion impacts in monolayer graphene \cite{zhang:2012,ojanpera:2014,zhao:2015,kononov:2021,vazquez:2021}, it may not necessarily damage the atomic structure.
The initially localized electronic excitations can quickly disperse both within and outside of the sample without transferring sufficient kinetic energy to individual atoms to overcome defect formation barriers.
For instance, the kinetic energy of emitted and captured electrons carries away 20\,--\,40\% of the energy initially transferred to graphene within the ion parameter range considered here \cite{vazquez:2021}, reducing the amount of energy remaining within the sample.
This figure depends on not only the number of emitted electrons, but also their energy spectrum, both of which in turn depend on ion energy and charge \cite{vazquez:2021}.

Furthermore, graphene has high carrier mobilities and weak electron-phonon coupling \cite{morozov:2008,chen:2008,bolotin:2008}, suggesting that valence electronic excitations within the material would delocalize too quickly to damage the atomic structure.
Accordingly, simulations of highly charged ions impacting a graphene layer represented as jellium \cite{gruber:2016,wilhelm:2018} predicted very large current densities which quickly spread electronic excitations throughout the material, preventing damage.
Nonetheless, experiments find large, nanoscale defects in few-layer carbon materials after irradiation by highly charged ions \cite{ritter:2013,hopster:2014,wilhelm:2015}, where localized electronic excitations are postulated to cause strong Coulombic repulsion of unscreened nuclei or weaken atomic bonds which then interact with the ambient environment.
Despite prior work, a characterization of how ion-induced electronic excitations transfer energy to individual atoms within a sample, thereby potentially producing defects, remains absent.

In addition to information relevant to damage processes, a comprehensive model of ion-beam microscopy must predict ion-induced electron emission, the quantity ultimately detected for imaging.
Comparatively little first-principles work exists in this space because of the high computational cost associated with the large supercells required \cite{kononov:2020,kononov:2021}.
Nonetheless, early work demonstrated the promise of first-principles methods for simulating electron emission in ion microscopy \cite{zhang:2012}.
Later, larger-scale calculations along with methodological advances \cite{kononov:2020} enabled predictions of the emitted electron yields detected in microscopy techniques \cite{kononov:2021}.
In particular, Ref.\ \onlinecite{kononov:2021} suggested that for proton-irradiated free-standing samples, exit-side (forward) electron emission may offer higher contrast than the traditionally detected entrance-side (backward) emission.
However, to our knowledge, no first-principles study has constructed simulated ion beam microscopy images using converged emitted electron yields or explained the physics underlying the dependence of image contrast on emission side or ion energy, mass, and charge.

Here, we extend prior work on first-principles simulations of ion-irradiated graphene by examining simulated microscopy images based on emitted electron yields calculated for a range of light-ion energies and impact points.
We also analyze the charge dynamics and atomic forces within the material in order to investigate the extent to which deposited energy remains localized near the impact point and may thus lead to defects.
Section \ref{sec:methods} describes our computational approach, Section \ref{sec:microscopy} discusses the simulated microscopy images, Section \ref{sec:damage} investigates charge dynamics and atomic forces within the graphene, and Section \ref{sec:conclusions} summarizes this contribution.

\section{Computational Methods}
\label{sec:methods}

The first-principles simulations were performed using real-time time-dependent density functional theory (TDDFT) \cite{runge:1984,marques:2004,ullrich:2012,ullrich:2014} as recently described in Refs.\ \onlinecite{kononov:2021,vazquez:2021}.
The initially ground-state graphene contained 112 carbon atoms, electron-ion interactions were described by HSCV pseudopotentials \cite{vanderbilt:1985}, exchange and correlation was treated with the adiabatic local density approximation \cite{zangwill:1980,zangwill:1981}, and the large supercells allowed Brillouin zone sampling using the $\Gamma$-point only.
A 100\,Ry plane-wave cutoff energy and a \SI{150}{\bohr} vacuum were previously found to achieve good convergence for electron emission in this system \cite{kononov:2021}.
A time step of \SI{1.0}{\atto\second} was used with the enforced time-reversal symmetry integrator \cite{castro:2004,draeger:2017}, which was previously shown to evolve similar systems accurately \cite{kang:2019,kononov:2020}.
Section \ref{sec:convergence} of the supplemental material demonstrates convergence with respect to the lateral supercell dimensions, i.e., the size of the graphene.
All TDDFT calculations were performed using the Qbox/Qb@ll code \cite{schleife:2012,draeger:2017}.

The charged projectile was inserted \SI{25}{\bohr} away from the graphene at the beginning of each simulation and proceeded along a normal trajectory with its velocity held constant.
A total of five different impact points were investigated as shown in \mbox{Fig.\ \ref{fig:traj_sampling}}, three of which sufficed to model atomic-resolution microscopy (see Sec.\ \ref{sec:sampling} of the supplemental material).
The cross-sectional supercell area of about \SI{3}{\nm^2} corresponds to a low ion dose of $3.4\times 10^{13}$\,cm$^{-2}$ within each single-impact simulation.
However, the ultimate simulated microscopy images constructed from multiple impact points in \mbox{Sec.\ \ref{sec:microscopy}} correspond to an effective ion dose of $3.0\times 10^{16}$\,cm$^{-2}$.
While the latter dose exceeds experimentally determined safe limits for helium ion microscopy of graphene \cite{fox:2013,hlawacek:2014}, nondestructive imaging of 2D samples at comparable total doses may still be possible with a different beam energy, under a beam current sufficiently low to allow healing between ion impacts, and/or through time-resolved measurements \cite{peng:2020}.
Alternatively, encapsulation can improve radiation hardness in some cases \cite{zan:2013,nanda:2015}.

Although helium ion microscopes employ a He$^+$ source, here we consider proton and He$^{2+}$ projectiles in order to avoid numerical challenges associated with accelerating a partially filled valence shell.
The potential energy contained in each light ion arising from ionization of the corresponding atom is 13.6, 24.6, and \SI{79}{\electronvolt} for H$^+$, He$^+$, and He$^{2+}$, respectively \cite{nist}.
Thus, we expect the effects of He$^+$ impacts to fall between the effects of proton and He$^{2+}$ impacts.
For fast ions ($\gtrsim 1$ atomic unit of velocity) that capture negligible electrons \cite{zhao:2015,kononov:2021,vazquez:2021} and essentially behave as classical point charges, the response to an He$^+$ ion will approach that of a proton.

Time-dependent electron densities calculated within TDDFT were analyzed according to the methods described in Refs.\ \onlinecite{kononov:2020,kononov:2021} to extract emitted electron yields from both sides of the material.
Briefly, the electron density was integrated in the entrance-side and exit-side vacuum regions, excluding the surface region within \SI{10.5}{\bohr} of the carbon atoms.
A dynamic boundary \cite{kononov:2021} allowed improved distinction between exit-side and entrance-side emissions in the presence of periodic boundary conditions.
The number of electrons captured by the passing ion was extracted from the density fitting technique introduced in Ref.\ \onlinecite{kononov:2020} using analytic H$^+$ orbitals and DFT orbitals calculated for an isolated He$^{2+}$ ion.
Captured electrons were then excluded from exit-side emission.
The results obtained for captured and emitted electrons represent expectation values and thus may have fractional values \cite{ullrich:2012}.

Notably, this work goes beyond the early simulations of helium ion microscopy in Ref.\ \onlinecite{zhang:2012} by using a very large vacuum region to achieve converged emitted electron yields.
Furthermore, we overcome earlier challenges in accounting for electron capture and separately analyze emission from both sides of the material.
We also consider a range of beam energies in order to guide optimal parameter selection, and our additional analysis of atomic forces allows quantitative comparisons related to defect formation processes.

\begin{figure*}[t]
    \centering
    \includegraphics{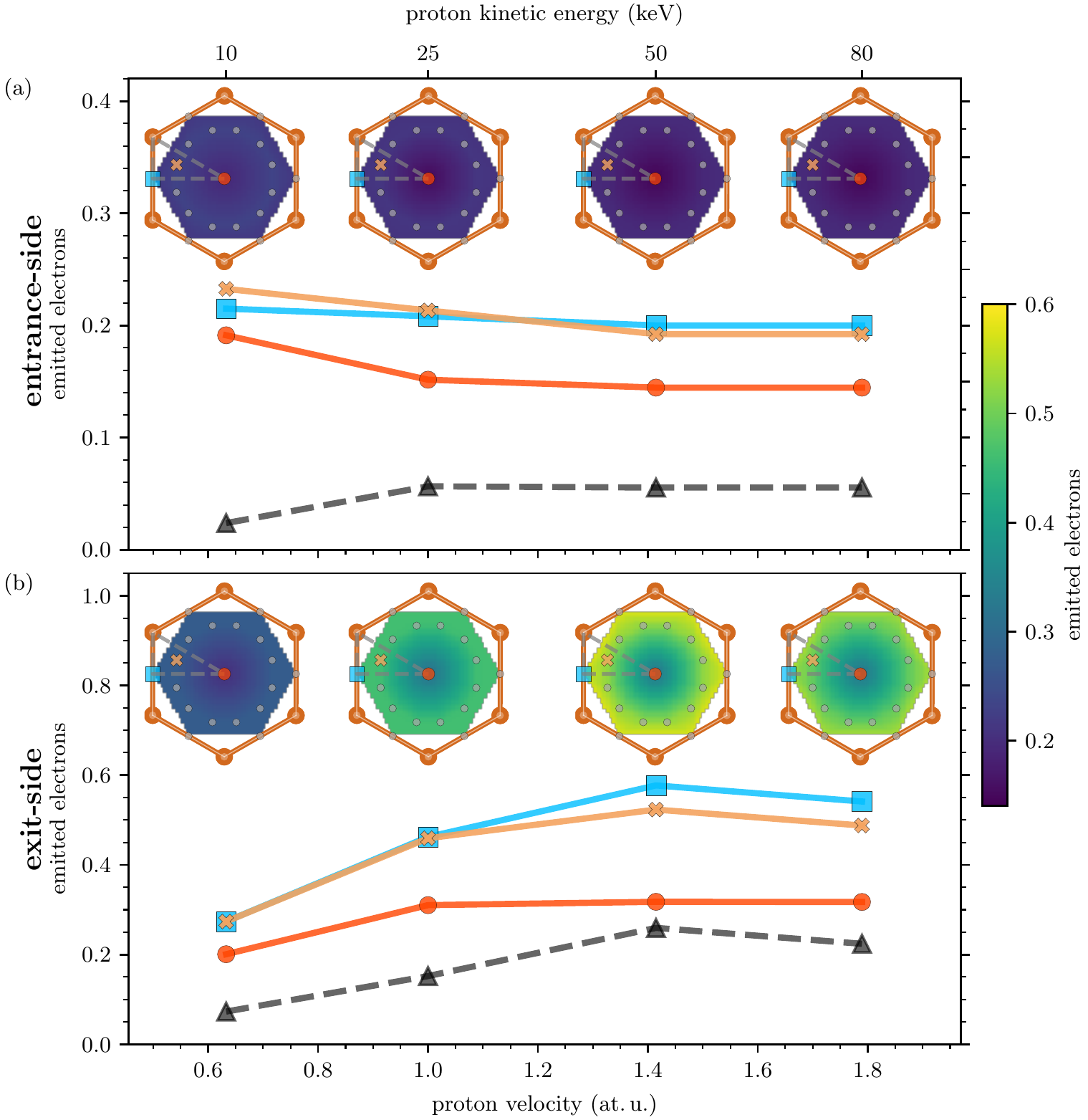}
    \caption{
    Emitted electron yields for different proton impact points and corresponding simulated microscopy images for (a) entrance-side and (b) exit-side emission.
    Colored symbols indicate explicitly simulated impact points within the gray, symmetry-irreducible triangle, and gray points indicate symmetry equivalents.
    Black triangles plot the contrast metric, defined in Eq.\ \eqref{eq:contrast} as the difference between emitted electron yields generated by impacts at the midpoint of a C\,--\,C bond (blue squares) and the center of a C ring (red circles).
    }
    \label{fig:proton_v}
\end{figure*}

\section{Simulated Microscopy Images}
\label{sec:microscopy}

\begin{figure*}[t!]
    \centering
    \includegraphics{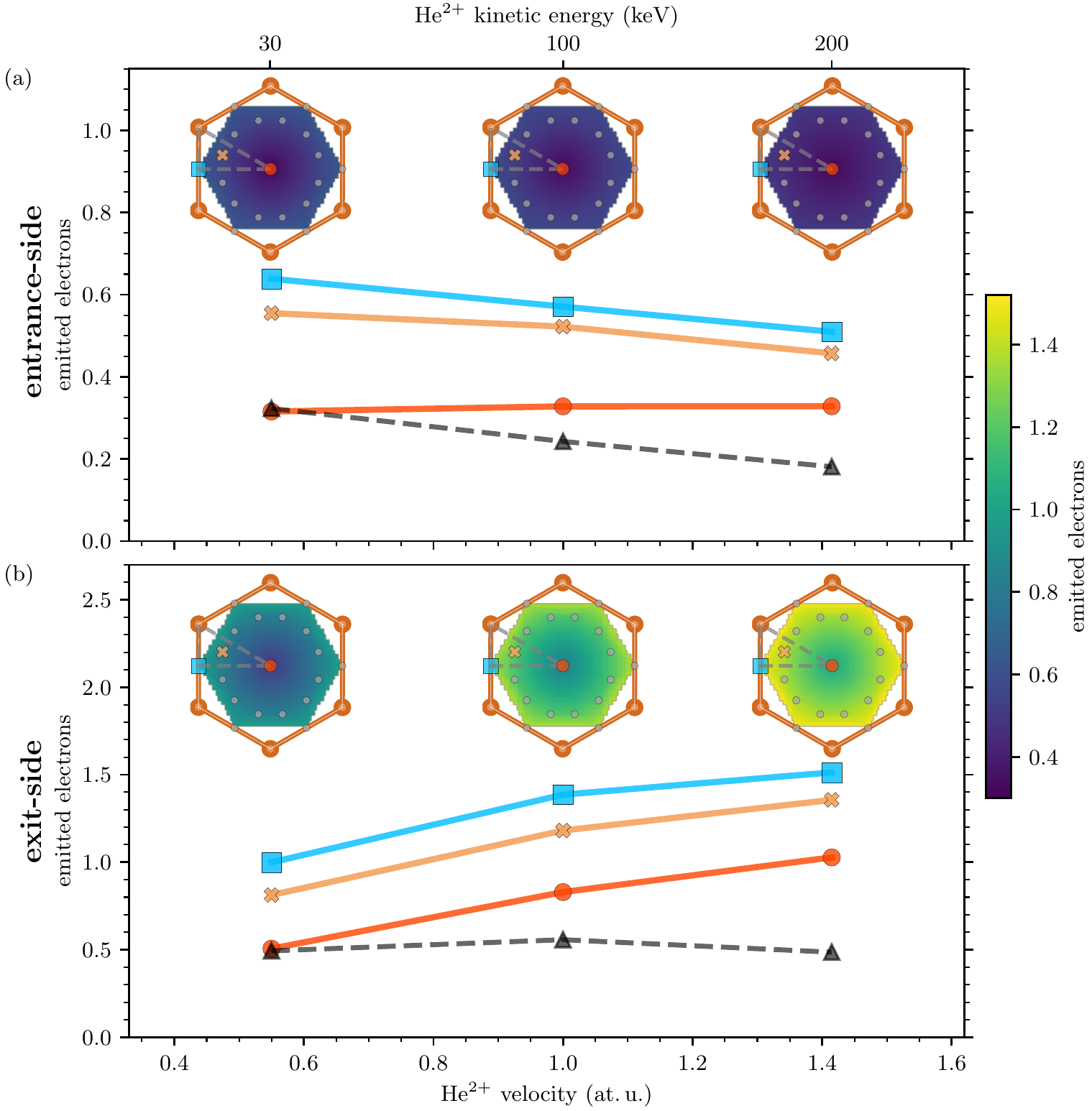}
    \caption{
    Emitted electron yields for different He$^{2+}$ impact points and corresponding simulated microscopy images for \mbox{(a) entrance-side} and (b) exit-side emission.
    Symbols are as indicated in Fig.\ \ref{fig:proton_v}.
    Note the different color bar scale from \mbox{Fig.\ \ref{fig:proton_v}}.
    }
    \label{fig:helium_v}
\end{figure*}

Microscopy techniques typically measure electron emission as the probe beam scans the sample, and the resulting map between beam position and observed emitted electron yield produces an image of the sample.
Analogously, simulated microscopy images can be generated by calculating emitted electron yields $\gamma$ for different ion impact points $\mathbf{x}$.
Here, we consider the total number of electrons emitted from either side of the material, i.e., our approach approximates
\begin{equation}
\gamma_j(\mathbf{x}) = \lim_{t\rightarrow\infty} \int_{V_j} n_{\mathbf{x}}(\mathbf{r},t)\; dr^3,
\label{eq:yield}
\end{equation}
where $j$ denotes entrance or exit side, $V_j$ is the corresponding vacuum region, and $n_{\mathbf{x}}$ is the electron density computed from TDDFT for the given ion parameters.
Practical limitations such as finite beam widths, detector collection efficiencies, and scan speeds introduce broadening factors and reduce the portion of this fundamental quantity that experiments ultimately measure.
We do not include these effects because they vary according to details of the experimental setup.

Since the pseudopotential approximation limits accuracy for impact points very close to the carbon atoms, we focus on a portion of the honeycomb lattice defined by the midpoints of the C\,--\,C bonds (see Fig.\ \ref{fig:proton_v}).
To reduce computational cost, we sample only a small, representative set of impact points and linearly interpolate between their symmetry equivalents.
In Sec.\ \ref{sec:sampling} of the supplemental material, we show that the three impact points illustrated in Fig.\ \ref{fig:proton_v} suffice to capture the essential features of simulated microscopy images produced using a larger data set.

Similar to the findings of Ref.\ \onlinecite{kononov:2021} and consistent with early experiments on thin foils \cite{mechbach:1975,rothard:1989,koschar:1989,ritzau:1998}, we generally predict stronger exit-side emission than entrance-side emission for both protons (see Fig.\ \ref{fig:proton_v}) and He$^{2+}$ ions (see \mbox{Fig.\ \ref{fig:helium_v}}) across all impact points.
This anisotropy manifests most dramatically for ion velocities $\geq 1$\,at.\,u., where protons and He$^{2+}$ ions produce 2\,--\,2.9 and 2.2\,--\,3.1 times more exit-side electrons than entrance-side electrons, respectively.
Assuming similar noise characteristics in detection of emitted electrons from both sides, which can be modeled with compound Poisson distributions \cite{timischl:2012,peng:2020}, measuring exit-side emission should allow 2\,--\,3 times lower beam doses to achieve a target signal to noise ratio.
A lower dose would reduce the likelihood of sample damage, improving prospects for nondestructive imaging.

In addition to a strong signal per ion impact, a successful ion-beam microscopy technique must achieve high contrast between distinct sample regions.
In particular, atomic resolution requires high sensitivity of electron emission to the ion's impact point within the lattice.
As a quantitative metric of contrast, we take the difference between emitted electron yields produced at the impact points closest to and furthest from the carbon atoms, i.e., the impact points at the midpoint of a C\,--\,C bond and at the center of a C ring:
\begin{equation}
    \mathrm{contrast} =
    \gamma\left(
    \begin{tikzpicture}[baseline=-0.65ex]
    \def\r{0.25}
    \def\pw{0.7pt}
    \draw [color=black, line width=0.5pt] (30:\r) -- (90:\r) -- (150:\r) -- (210:\r) -- (270:\r) -- (330:\r) -- (30:\r);
    \node at (30:\r) [circle,fill,inner sep=\pw,color=black]{};
    \node at (90:\r) [circle,fill,inner sep=\pw,color=black]{};
    \node at (150:\r) [circle,fill,inner sep=\pw,color=black]{};
    \node at (210:\r) [circle,fill,inner sep=\pw,color=black]{};
    \node at (270:\r) [circle,fill,inner sep=\pw,color=black]{};
    \node at (330:\r) [circle,fill,inner sep=\pw,color=black]{};
    \node at (180:{\r * sqrt(3)/2}) [circle,fill,color=red, inner sep=\pw]{};
    \end{tikzpicture}
    \right) - \gamma\left(
    \begin{tikzpicture}[baseline=-0.65ex]
    \def\r{0.25}
    \def\pw{0.7pt}
    \draw [color=black, line width=0.5pt] (30:\r) -- (90:\r) -- (150:\r) -- (210:\r) -- (270:\r) -- (330:\r) -- (30:\r);
    \node at (30:\r) [circle,fill,inner sep=\pw,color=black]{};
    \node at (90:\r) [circle,fill,inner sep=0\pw,color=black]{};
    \node at (150:\r) [circle,fill,inner sep=\pw,color=black]{};
    \node at (210:\r) [circle,fill,inner sep=\pw,color=black]{};
    \node at (270:\r) [circle,fill,inner sep=\pw,color=black]{};
    \node at (330:\r) [circle,fill,inner sep=\pw,color=black]{};
    \node at (0:0) [circle,fill,color=red, inner sep=\pw]{};
    \end{tikzpicture}
    \right),
    \label{eq:contrast}
\end{equation}
where the red point indicates the ion's impact point relative to the carbon lattice.

As suggested by Ref.\ \onlinecite{kononov:2021}, we find that exit-side electron emission indeed produces higher contrast than entrance-side emission for both protons (see Fig.\ \ref{fig:proton_v}) and He$^{2+}$ ions (see Fig.\ \ref{fig:helium_v}).
This trend holds across the entire ion velocity range considered here, with exit-side contrast exceeding entrance-side by a factor of 2.7\,--\,4.7 for protons and 1.5\,--\,2.7 for He$^{2+}$ ions.
Fig.\ \ref{fig:proton_v}a shows that for protons, entrance-side contrast remains below 0.06 even for a \SI{10}{\kilo\electronvolt} beam, where we predict maximum entrance-side emission.
On the other hand, in Fig.\ \ref{fig:proton_v}b, exit-side contrast achieves a much higher maximum of 0.26 for \SI{50}{\kilo\electronvolt} protons.
This proton energy maximizes exit-side electron emission for impact points near the carbon atoms, while electron emission induced by protons impacting at the center of a carbon ring is not as sensitive to proton energy.

The \SI{30}{\kilo\electronvolt} beam energy commonly used in helium ion microscopes offers the highest entrance-side contrast of 0.32 among the three He$^{2+}$ ion energies presented in \mbox{Fig.\ \ref{fig:helium_v}a}.
This energy both maximizes entrance-side electron emission near carbon atoms and minimizes entrance-side emission at the center of a carbon ring.
The highest exit-side contrast of 0.56 is instead achieved by \SI{100}{\kilo\electronvolt} He$^{2+}$ ions, though the other two He$^{2+}$ energies also produce relatively high contrast metrics of 0.49 (see \mbox{Fig.\ \ref{fig:helium_v}b)}.
Generally, we predict that a higher beam energy is needed to optimize image contrast achieved by detecting exit-side emitted electrons than entrance-side emitted electrons.
Beam energy also affects the rate of energy deposition in the sample and damage processes, which we further examine in \mbox{Sec.\ \ref{sec:damage}}.

The differing trends of entrance-side and exit-side emission may derive from distinct physical mechanisms contributing to emission from either side.
Kinetic emission, where the impacting ion directly transfers momentum and kinetic energy to individual electrons, forms one such mechanism and is commonly modeled as proportional to electronic stopping power \cite{sternglass:1957,baragiola:1979,koschar:1989,ramachandra:2009}.
Electrons excited in this way are highly anisotropic, preferentially traveling in the same direction as the projectile because of momentum conservation \cite{dehaes:1993}.
In a thick sample, these initially excited electrons would experience further scattering which could eventually allow them to escape from the entrance-side surface.
However, since transverse electron mean free paths in graphene are comparable to the layer thickness \cite{geelen:2019}, many of these energetic electrons do not undergo enough collisions to reverse their momentum and thus simply escape from the exit-side surface.

Another mechanism, so-called potential emission, instead relies on projectile neutralization processes releasing potential energy and exciting electrons through Auger transitions or autoionization \cite{aumayr:2007,schwestka:2019}.
Potential emission occurs as the projectile approaches the material, i.e., on the entrance side, and dominates for slow, highly charged ions \cite{kurz:1992,kurz:1993,aumayr:1993}.
In principle, light ions could also induce potential emission since the potential energies stored in H$^+$ and He$^{2+}$ ions (13.6 and \SI{79}{\electronvolt}, respectively \cite{nist}) exceed twice the graphene work function of \SI{4.6}{\electronvolt} \cite{yan:2012}, the minimum required for Auger neutralization.
However, Ref.\ \onlinecite{kononov:2021} predicted negligible electron emission from both sides of graphene after impact by protons below a threshold velocity of 0.1\,--\,0.2 atomic units required to kinetically excite electrons over the work function.
Since potential emission would still occur below this kinetic emission threshold, we conclude that potential emission does not contribute significantly here.

Finally, electron emission can result from the decay of plasmons excited through either kinetic or potential energy transfer processes \cite{rosler:1992,baragiola:2007}.
Experimental characterization of this mechanism has proved elusive because of challenges in isolating it from other electron emission processes \cite{riccardi:2003,bajales:2005}.
Nonetheless, the contribution of plasmon-assisted electron emission has been predicted to depend weakly on ion velocity beyond its peak \cite{rosler:1992}, similar to the velocity-dependence found for entrance-side emission in this work and in Ref.\ \onlinecite{kononov:2021}.

So, we propose that entrance-side electron emission from a 2D material irradiated by light ions mainly arises from plasmon decay, explaining its lower sensitivity to ion energy compared to both exit-side emission and electronic stopping power.
On the other hand, exit-side electron emission also contains a contribution from kinetic emission and thus follows a similar velocity-dependence as electronic stopping power, which features a prominent peak for \mbox{50\,--\,80\,keV} protons impacting graphene \cite{kononov:2021}.

Additionally, the anisotropic nature of kinetic emission from 2D materials explains the higher contrast predicted for exit-side microscopy.
The electron density near the impact point can be expected to strongly influence cross sections for the binary collisions leading to kinetic emission.
Meanwhile, plasmon excitation occurs over a longer length scale and therefore is not as sensitive to ion impact point, resulting in lower contrast for the primarily plasmon-mediated entrance-side emission.
We note, however, that TDDFT with an adiabatic exchange-correlation functional may not be capable of accurately capturing complex processes such as plasmon decay and Auger transitions.
More work is needed to confirm the role of these mechanisms and characterize the limitations of the theoretical approach in describing them.

\section{Damage Indicators}
\label{sec:damage}

In addition to producing strong electron emission that is highly sensitive to the ion impact point, an ideal imaging technique should also avoid disturbing the atomic structure of the material.
Our first-principles approach allows direct analysis of charge dynamics and atomic forces in ion-irradiated graphene, offering detailed information about the early stages of any damage processes.
Here, we focus on ions impacting along a C\,--\,C bond because among the impact points considered in this work, we expect these to be most likely to damage atomic structure.
We report additional results for other impact points in Sec.\ \ref{sec:extra_forces} of the Supplemental Material.
We note that impact points even closer to a C atom, especially head-on collisions as studied in Ref.\ \onlinecite{kretschmer:2022}, can cause damage through ion-ion scattering, but in this work we specifically investigate the possibility of damage mechanisms arising from electronic excitations.

\begin{figure}
    \centering
    \includegraphics{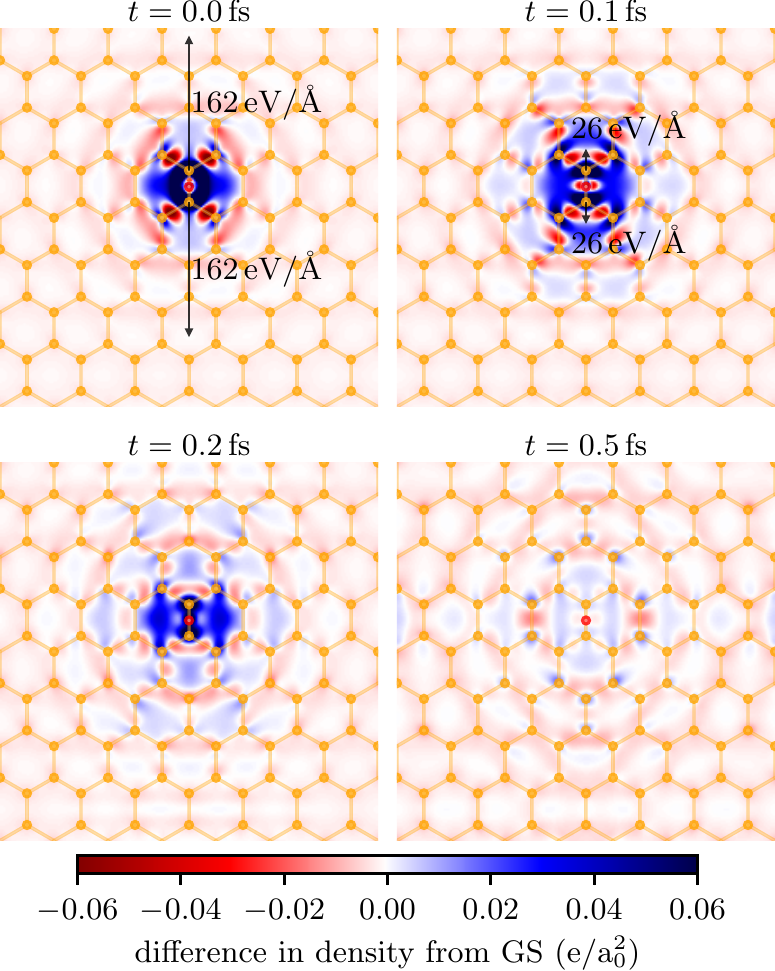}
    \caption{Snapshots of the charge distribution in graphene (orange) after a  \SI{30}{\kilo\electronvolt} He$^{2+}$ ion impacts at the midpoint of a C\,--\,C bond (red point).
    Red (blue) regions indicate lower (higher) electron density relative to the initial ground state, where the electron density has been integrated along the out-of-plane direction over a \SI{21}{\bohr}-thick slab centered on the graphene plane.
    In-plane atomic forces above \SI{10}{\electronvolt\per\angstrom} are indicated by black arrows.
    See web version for a video of the full time evolution.
    }
    \label{fig:charge_evolution}
\end{figure}

We find that the charge induced in graphene by light ions quickly spreads out and equilibrates within the few-femtosecond simulations (see \mbox{Fig.\ \ref{fig:charge_evolution}}), indicating that light ions are indeed unlikely to damage the atomic structure of graphene.
To explicitly analyze the charge dynamics in the graphene, we compute the total charge excited out of the initial ground state as
\begin{equation}
    N(t) = \int_{V_G} |n(\mathbf{r},t) - n(\mathbf{r},0)|\; dr^3,
    \label{eq:charge_deviation}
\end{equation}
where $V_G$ denotes a \SI{21}{\bohr}-thick slab containing the graphene and its surface regions. 
Positively and negatively charged contributions to $N(t)$ can be analogously defined as integrals over portions of $V_G$ where $n(\mathbf{r},t) < n(\mathbf{r},0)$ and $n(\mathbf{r},t) > n(\mathbf{r},0)$, respectively.
These contributions represent the number of excited holes and excited electrons, which may occupy complicated, time-dependent volumes.
This density-based method falters when excited electrons and holes overlap within the analyzing volume, but its results agree qualitatively with an analysis of ground-state Kohn-Sham orbital occupations \cite{kononov:2021} (see \mbox{Fig.\ \ref{fig:excited_es_vs_KS}} in the Supplemental Material).

As shown in Figs.\ \ref{fig:excited_es}a and b, the numbers of excited electrons and holes increase symmetrically before impact, reflecting initial polarization of the sheet and subsequent excitation of plasmons.
After impact, the number of excited electrons decays while the number of excited holes remains roughly constant as electron emission and capture by the ion remove electrons and the graphene equilibrates to a net positive charge distributed across the entire sheet.
Interestingly, Fig.\ \ref{fig:excited_es}c shows that the total excited charge depends strongly on ion charge, but not ion energy: at low energies, higher charge capture \cite{kononov:2021} compensates for weaker electron emission, resulting in similar strength excitations in the graphene.
Both before and after impact, the amount of excited charge gently oscillates over time, and these fluctuations become more apparent for a thinner analyzing slab (see Fig.\ \ref{fig:excited_es_supp} in the Supplemental Material).
Their period of around \SI{0.3}{\femto\second} roughly corresponds to the \SI{14.6}{\electronvolt} plasmon mode in graphene \cite{eberlein:2008}, supporting the notion that plasmonic excitations play an important role in this system.

We also estimate a charge equilibration time scale by fitting $N(t)$ to an exponential decay model 
\begin{equation}
    N(t) \approx (N(0)-N_\infty)\, e^{-t/\tau} + N_\infty
    \label{eq:fit}
\end{equation}
for times after the projectile has traveled at least \SI{5}{\bohr} away from the analyzing volume.
This cutoff distance mitigates interference of electrons captured by the ion with the analysis.
We find characteristic equilibration times $\tau\approx0.1$\,---\,\SI{0.3}{\femto\second} and note that this quantity is difficult to converge with respect to the graphene dimensions and additionally varies with analyzing slab thickness and cutoff distance (see Sec.\ \ref{sec:convergence} and \ref{sec:extra_charge} of the Supplemental Material for more details).
Ion energy does not significantly influence equilibration times within the uncertainties of this analysis, but we consistently find that equilibration times after a He$^{2+}$ impact are about 1.3 times longer than after a proton impact.

\begin{figure}
    \centering
    \includegraphics{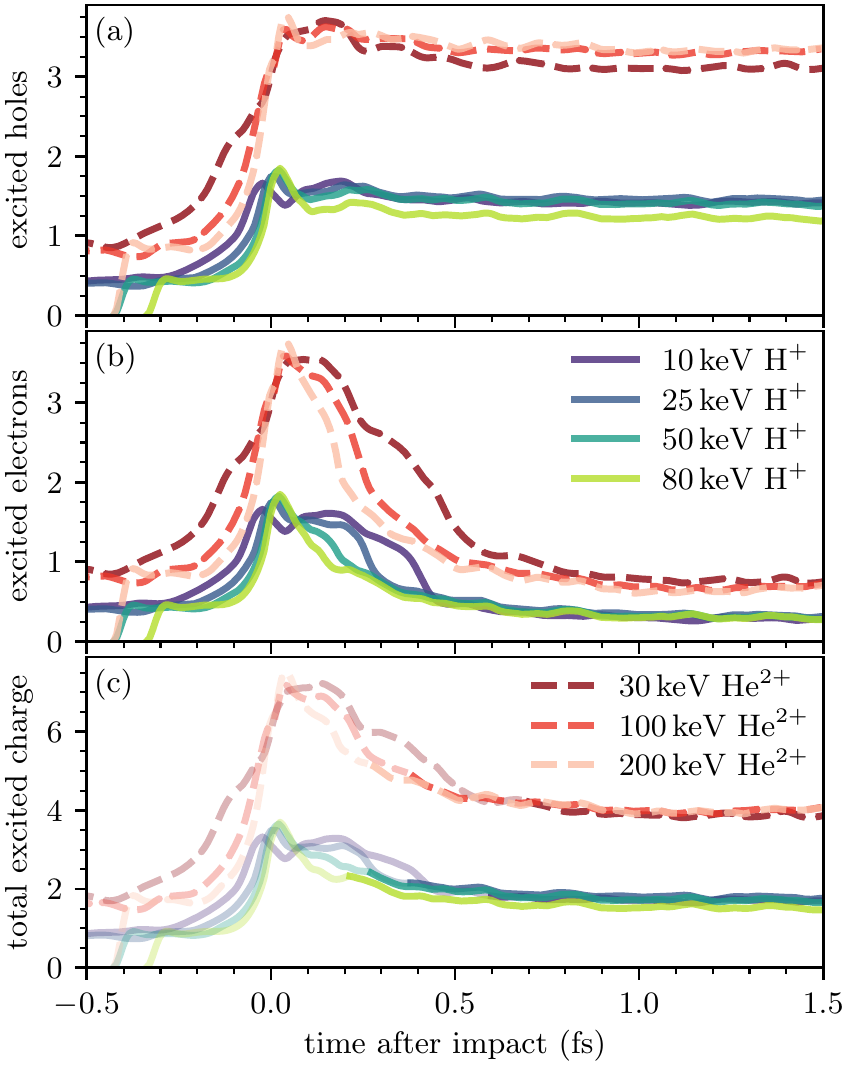}
    \caption{
    The number of excited holes (a), the number of excited electrons (b), and the total excited charge (c) in the irradiated graphene, as defined by Eq.\ \eqref{eq:charge_deviation} and accompanying text.
    In (c), light (dark) curves indicate data before (after) the cutoff time used in fitting to Eq.\ \eqref{eq:fit} for estimating the charge equilibration time.
    }
    \label{fig:excited_es}
\end{figure}

\begin{figure}
    \centering
    \includegraphics{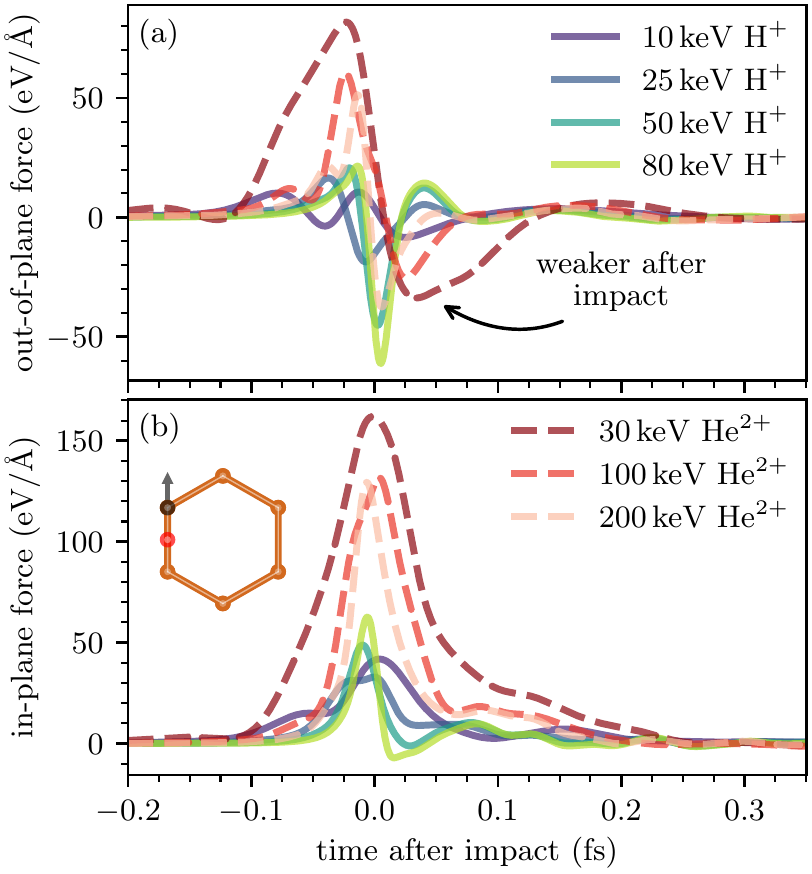}
    \caption{Instantaneous out-of-plane (a) and in-plane (b) force experienced by a nearest C atom after an ion impacts at the midpoint of a C-C bond.
    Positive out-of-plane forces point in the direction of projectile motion, and positive in-plane forces point away from the impact point.
    The inset illustrates the geometry, with the ion impact point shown in red and the carbon atom in question shown in black.
    }
    \label{fig:forcesF}
\end{figure}

Because of the ultrafast charge equilibration, carbon atoms near the impact point only experience large-magnitude forces on the order of \SI{100}{\electronvolt\per\angstrom} during a sub-fs interval around the time of impact (see Fig.\ \ref{fig:forcesF}).
The magnitude of the Hellmann-Feynman force acting on one of the nearest carbon atoms decays to less than 3\% of its maximum within \SI{0.35}{\femto\second} of impact.

\begin{figure}
    \centering
    \includegraphics{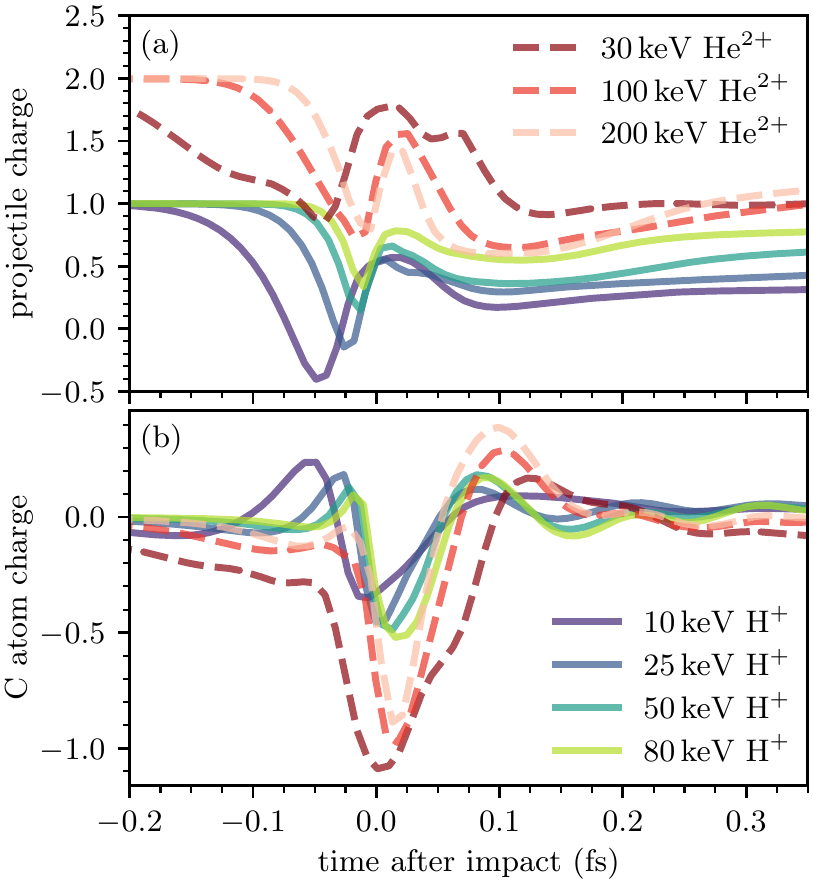}
    \caption{Instantaneous DDEC6 \cite{manz:2016,gabaldon:2016} charge computed for (a) the projectile and (b) the nearest C atom after an ion impacts at the midpoint of a C-C bond.
    }
    \label{fig:ddec6F}
\end{figure}

Beyond their time scale, the dynamical evolution of the atomic forces is itself interesting.
During proton impacts, the instantaneous out-of-plane force acting on the nearest C atom (see Fig.\ \ref{fig:forcesF}a) shows complex oscillatory behavior 0.05\,--\,\SI{0.09}{\femto\second} in period.
The corresponding oscillation energies of 45\,--\,\SI{80}{\electronvolt} are too high to be explained by 4.7 and \SI{14.6}{\electronvolt} plasmon modes in graphene \cite{eberlein:2008}.
They may instead arise from a complex interplay of charge dynamics within the material and projectile charge capture processes dynamically modifying screening of the Coulombic repulsion by the incident ion.
Somewhat lower frequency oscillations of 30\,--\,\SI{40}{\electronvolt} appear in the effective charges of the projectile and nearest C atom, as computed with the DDEC6 charge partitioning method \cite{manz:2016,gabaldon:2016} in \mbox{Fig.\ \ref{fig:ddec6F}.}

The out-of-plane force induced by He$^{2+}$ ions, on the other hand, generally points away from the projectile, i.e., maintains positive values before impact and negative values after impact (see Fig.\ \ref{fig:forcesF}a).
Meanwhile, the instantaneous in-plane forces plotted in Fig.\ \ref{fig:forcesF}b remain largely positive (i.e., almost always point away from the impact) across all projectiles.
This behavior would be expected for partially screened Coulombic repulsion between a partially neutralized projectile and partially ionized carbon atom.
The weaker out-of-plane repulsion after impact may then be attributed to captured electrons further screening the projectile's charge.
However, the behavior of the electron density does not necessarily support this simple picture: during impact, the incident ion attracts additional electrons to a region spanning multiple lattice constants (see Fig.\ \ref{fig:charge_evolution}).
As shown in Fig.\ \ref{fig:ddec6F}, this excess electron density results in a negative effective charge on the nearest carbon atom, which would instead lead to attractive forces.
In reality, the non-equilibrium charge distribution also extends to other nearby carbon atoms, which also contribute to the net atomic forces close to the impact point.
Moreover, the complex density perturbations cannot be unequivocally allocated to individual atomic charges, precluding a simple electrostatic interpretation of the dynamic atomic forces.

If sufficiently large, the momentum transferred to carbon atoms through the ultrashort force pulses of Fig.\ \ref{fig:forcesF} may lead to defect formation.
We calculate the momentum transferred to the carbon atom closest to the impact point by evaluating the impulse
\begin{equation}
    \mathbf{I} = \int_{t_0}^{t_1} \mathbf{F}(t) \; dt,
\label{eq:impulse}
\end{equation}
where $\mathbf{F}(t)$ is the time-dependent Hellmann-Feynman force on the carbon nucleus, $t_0<-0.25$\,fs is near the beginning of the TDDFT simulation \footnote{We exclude $\sim$\SI{0.09}{\femto\second} at the beginning of the TDDFT simulations from the impulse calculations because of transient fictitious forces on the order of 1\,--\,\SI{2}{\electronvolt\per\angstrom} arising from sudden insertion of the projectile.}, and $t_1>1.4$\,fs is at the end.
Given initially motionless carbon atoms, $|I|^2/(2M)$ then gives the kinetic energy transferred to the given atom, where $M$ is its mass.

We find that the net impulse delivered to these nearest carbon atoms is quite small, corresponding to a kinetic energy transfer of at most \SI{0.14}{\electronvolt} in the case of a \SI{30}{\kilo\electronvolt} He$^{2+}$ ion, 
a small fraction of the nearly \SI{100}{\electronvolt} total energy deposited by this ion.
For the most promising microscopy candidate as identified in Sec.\ \ref{sec:microscopy}, a \SI{100}{\kilo\electronvolt} He$^{2+}$ ion, the net impulse is only \SI{40}{\milli\electronvolt}, and the largest net impulse delivered by a proton is only \SI{8.5}{\milli\electronvolt}.
These values depend strongly on impact parameter, with protons and He$^{2+}$ ions impacting at the center of a carbon ring delivering at most \SI{0.2}{\milli\electronvolt} and \SI{2.3}{\milli\electronvolt} to one of the nearest carbon atoms, respectively.
The net out-of-plane impulse is always aligned with the projectile's momentum and decreases with increasing ion energy, a trend consistent with decaying nuclear stopping power within the energy regime presently considered \cite{ziegler:2010}.
Much stronger instantaneous in-plane forces than corresponding out-of-plane forces (see Fig.\ \ref{fig:forcesF}) lead to in-plane impulses that exceed out-of-plane impulses by a factor of $\sim$5\,--\,20.
In-plane impulses also decrease with increasing ion energy, but not as quickly as out-of-plane impulses, meaning that disturbances produced by faster ions are increasingly directed in-plane.

These kinetic energy transfers are much smaller than point defect formation energies of 5\,--\,\SI{8}{\electronvolt} in graphene \cite{banhart:2011,krasheninnikov:2006,ma:2009,faccio:2010,ozcelik:2013} and smaller still than the bond rotation energy barrier of about \SI{10}{\electronvolt} \cite{li:2005,wang:2012} and the displacement threshold energy of about \SI{20}{\electronvolt} \cite{yazyev:2007,wang:2012,merrill:2015}.
Thus, we have shown that electronic excitations due to single impacts by light ions are not likely to introduce defects.
Instead, the primary immediate effect on the carbon lattice is stretching impacted C-C bonds.

For multiple nearby ion impacts under realistic imaging conditions, in-plane impulses should approximately cancel out because of symmetry considerations.
Based on the largest out-of-plane impulse of \SI{6}{\milli\electronvolt} predicted in this work for a \SI{30}{\kilo\electronvolt} He$^{2+}$ ion, the bonds of a carbon atom would need to be impacted around $10^3$ times before that atom would acquire enough kinetic energy to exceed defect formation energies.
The corresponding ion dose would be on the order of $10^{19}$\,cm$^{-2}$, far beyond the $\sim$\,$10^{14}$\,cm$^{-2}$ dose of \SI{30}{\kilo\electronvolt} He$^+$ ions above which experiments observed damage in graphene \cite{fox:2013}.
Although average nuclear stopping is much smaller than electronic stopping in this regime \cite{ziegler:2010}, the probability of at least one very close collision grows with ion dose.
Since head-on collisions have been shown to eject carbon atoms from graphene for ion energies above tens of eV \cite{kretschmer:2022}, we conclude that damage in graphene under high-dose light-ion irradiation results from ion-ion scattering rather than electronic excitations.

While the carbon atom closest to the impact point experiences the largest forces and impulses, the impinging ion also perturbs other nearby atoms.
We evaluate the total momentum transferred to the entire graphene sheet by summing the net out-of-plane impulses on each atom:
\begin{equation}
    I_{\mathrm{tot}} = \sum_j \int_{t_0}^{t_1} F_j(t) \; dt,
\label{eq:tot_impulse}
\end{equation}
where $F_j$ are out-of-plane forces on individual carbon atoms.
We find that $I_{\mathrm{tot}}$ accounts for about 45\,--\,90\% (35\,--\,70\%) of the total momentum that would be lost by a He$^{2+}$ (proton) projectile, with the remainder of the momentum transferred to emitted and captured electrons.
The largest total momentum transfer of \SI{0.066}{\amu\angstrom\per\femto\second}
again occurs for a \SI{30}{\kilo\electronvolt} He$^{2+}$ ion.
Similar to the trend noted earlier for impulses experienced by individual carbon atoms, the total impulse decreases with increasing ion energy, with 0.039 and \SI{0.024}{\amu\angstrom\per\femto\second} delivered by \SI{100}{\kilo\electronvolt} and \SI{200}{\kilo\electronvolt} He$^{2+}$ ions, respectively.
The total impulse delivered by protons is much smaller, ranging from 0.006 to \SI{0.014}{\amu\angstrom\per\femto\second}.

These momentum transfers can be compared to the adhesion energy of graphene on its support grid to assess the possibility of sample detachment during imaging.
The adhesion energy of graphene on copper, a common material for transmission electron microscopy (TEM) grids, is about \SI{0.7}{\joule\per\meter^2} \cite{yoon:2012,xin:2017}.
For the case of a very fine grid with \SI{7.5}{\micro\meter} holes and \SI{5}{\micro\meter} supports (as used for TEM in Ref.\ \onlinecite{ackerman:2016}), we estimate that at least $1.2\times10^{10}$ single impacts of \SI{30}{\kilo\electronvolt} He$^{2+}$ or equivalently, an ion dose of at least $7.7\times 10^{15}$\,cm$^{-2}$ would be needed to transfer enough total momentum to overcome the adhesion energy.
The lower impulses delivered by \SI{100}{\kilo\electronvolt} and \SI{200}{\kilo\electronvolt} He$^{2+}$ ions would allow higher doses of at least $1.3\times 10^{16}$\,cm$^{-2}$ and $2.1\times 10^{16}$\,cm$^{-2}$, respectively.
For protons, this estimated dose limit ranges from 3.6 to $8.0\times 10^{16}$\,cm$^{-2}$.
However, realistic beam currents on the order of 1\,--\,\SI{10}{\pico\ampere} \cite{ward:2006,fox:2013} would require several minutes of imaging to apply such doses.
During this time, impulses induced by individual ion impacts would dissipate and decohere into phonon modes.
Therefore, these dose estimates represent lower bounds on the dose at which sample detachment could occur.

\section{Conclusions}
\label{sec:conclusions}

We predict that detecting exit-side electron emission after light-ion irradiation would produce higher contrast images of suspended graphene than existing ion microscopy techniques relying on entrance-side electron emission.
Somewhat higher beam energies of 50\,--\,\SI{100}{\kilo\electronvolt} achieve maximal contrast in exit-side emission than typically used in helium ion microscopy.
These more energetic ions deposit less energy into the nuclear subsystem, likely leading to less damage to the atomic structure at the same ion dose.
Much stronger exit-side electron emission compared to entrance-side emission could allow lower ion doses without sacrificing image brightness, further reducing damage to the sample.

We also find that the charge induced in graphene by single light-ion impacts dissipates on a sub-fs timescale, indicating that deposited energy does not remain localized long enough to generate defects in the atomic structure.
Carbon atoms near the impact likewise experience large forces only during a sub-fs period, only gaining small kinetic energies on the order of \SI{0.1}{\electronvolt} or less.
This energy transfer is far too small to overcome defect formation barriers, but may deform bonds.

This work offers practical insights for advancing ion beam techniques for nondestructive imaging of thin materials.
Experimental work is needed to confirm these predictions, and further theoretical work may investigate damage processes caused by electronic excitations induced by higher charge ions.\\

\begin{acknowledgments}
We gratefully acknowledge helpful discussions with Ed Bielejec, Michael Titze, and Gyorgy Vizkelethy.
AK, AO, and ADB were partially supported by the US Department of Energy Science Campaign 1.
This material is based upon work supported by the National Science Foundation under Grant OAC-1740219.
Support from the IAEA F11020 CRP ``Ion Beam Induced Spatiotemporal Structural Evolution of Materials: Accelerators for a New Technology Era" is gratefully acknowledged.

This research is part of the Blue Waters sustained-petascale computing project, which is supported by the National Science Foundation (awards OCI-0725070 and ACI-1238993) and the state of Illinois.
Blue Waters is a joint effort of the University of Illinois at Urbana-Champaign and its National Center for Supercomputing Applications.
This work made use of the Illinois Campus Cluster, a computing resource that is operated by the Illinois Campus Cluster Program (ICCP) in conjunction with the National Center for Supercomputing Applications (NCSA) and which is supported by funds from the University of Illinois at Urbana-Champaign.

Sandia National Laboratories is a multi-mission laboratory managed and operated by National Technology and Engineering Solutions of Sandia, LLC, a wholly owned subsidiary of Honeywell International, Inc., for DOE's National Nuclear Security Administration under contract DE-NA0003525.
This paper describes objective technical results and analysis.
Any subjective views or opinions that might be expressed in the paper do not necessarily represent the views of the U.S. Department of Energy or the United States Government.
\end{acknowledgments}

\bibliography{main}

\clearpage
\onecolumngrid
\begin{center}
\textbf{\large Supplemental Materials: First-principles simulation of light-ion microscopy of graphene}
\end{center}

\setcounter{page}{1}
\setcounter{section}{0}
\setcounter{figure}{0}
\setcounter{equation}{0}
\renewcommand{\thesection}{S\arabic{section}}
\renewcommand{\theHsection}{S\thesection}
\renewcommand{\thefigure}{S\arabic{figure}}
\renewcommand{\theHfigure}{S\thefigure}
\renewcommand{\theequation}{S\arabic{equation}}

\twocolumngrid
\section{Convergence of graphene dimensions}
\label{sec:convergence}

To evaluate finite-size effects arising from artificial interactions among periodic images of the supercell, we simulated \SI{25}{\kilo\electronvolt} protons impacting at the center of a carbon ring for graphene supercells of different dimensions.
We compared results obtained for the \SI{2.9}{\nano\meter^2}, 112-atom sheet used in the main text to those calculated for a \SI{1.6}{\nano\meter^2}, 60-atom sheet and a \SI{6.3}{\nano\meter^2}, 240-atom sheet.
To reduce computational cost, a shorter vacuum of \SI{100}{\bohr} was used for these tests.

We found excellent convergence of total energy deposition, with only 1\% differences among the different graphene supercells.
The number of electrons captured by the proton does not depend significantly on the supercell size, but the emitted electron yields are more sensitive, with a 3\% (5\%) difference between the larger (smaller) two graphene supercells.
The out-of-plane impulse on the nearest carbon atom is also well-converged, with a 2\% (5\%) difference between the larger (smaller) two supercells.
The total out-of-plane impulse on the entire graphene sheet changes by 3\% (7\%) between the larger (smaller) two supercells.
Thus, we consider the 112-atom graphene supercell to achieve acceptable convergence at a feasible computational cost.

\begin{figure}[h]
    \centering
    \includegraphics{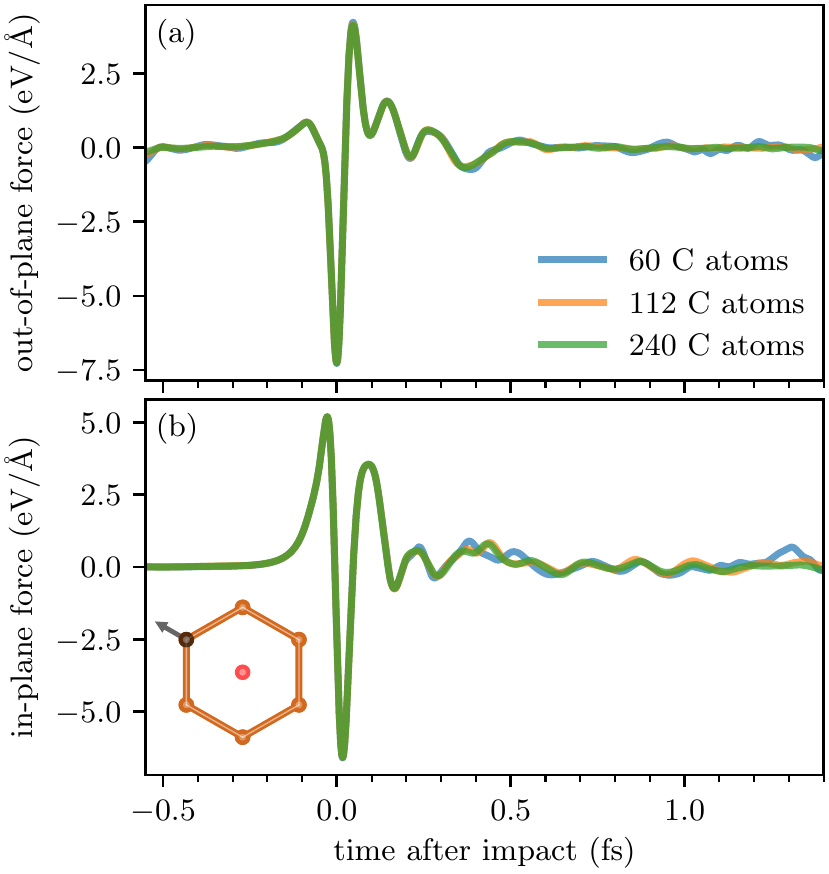}
    \caption{
    Instantaneous out-of-plane (a) and in-plane (b) force experienced by a nearest C atom after a \SI{25}{\kilo\electronvolt} proton impacts as shown in the inset.
    Results using different size graphene supercells are compared.
    Positive out-of-plane forces point in the direction of projectile motion, and positive in-plane forces point away from the impact point.
    }
    \label{fig:forcesA_conv}
\end{figure}

\begin{figure}[b]
    \centering
    \includegraphics{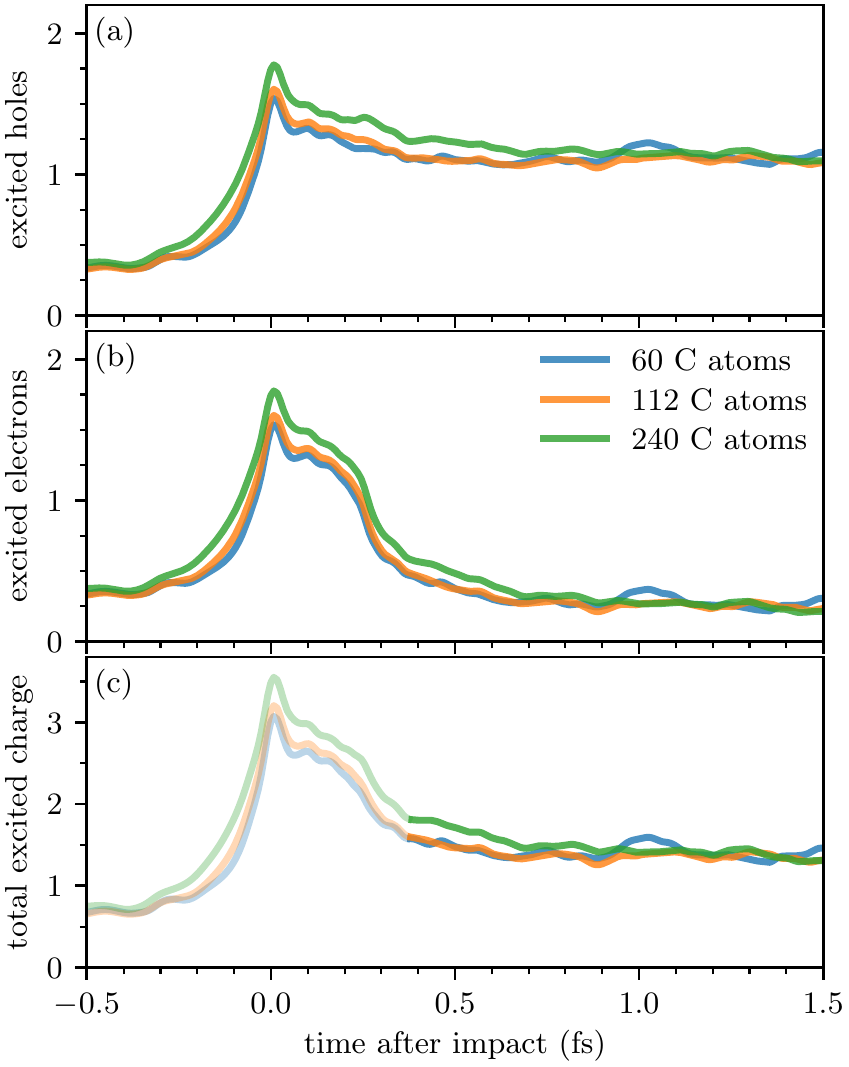}
    \caption{
    The number of excited holes (a), the number of excited electrons (b), and the total excited charge (c) in the irradiated graphene, as defined by Eq.\ \eqref{eq:charge_deviation} and accompanying text.
    Results using different size graphene supercells are compared for the case of a \SI{25}{\kilo\electronvolt} proton impacting at the center of a carbon ring.
    In (c), light (dark) curves indicate data before (after) the cutoff time used in fitting to Eq.\ \eqref{eq:fit} for estimating the charge equilibration time.
    }
    \label{fig:excited_es_conv}
\end{figure}

However, artificial in-plane charge dynamics across periodic boundaries still cause some finite-size effects in the in-plane atomic forces and charge equilibration analysis.
While the instantaneous force on the nearest carbon atom agrees very closely across the three different graphene supercells before and during impact, small deviations begin to appear about \SI{0.4}{\femto\second} after impact (see Fig.\ \ref{fig:forcesA_conv}).
This leads to comparatively slow convergence of the net in-plane impulse on the nearest carbon atom: we find a 10\% (13\%) difference between the larger (smaller) two supercells.
Furthermore, the charge equilibration time scale extracted for the largest supercell is 1.5\,--\,1.7 times longer than for the two smaller supercells.
Failure to converge this quantity despite similar qualitative evolution of the excited charge data from which it is derived (see Fig.\ \ref{fig:excited_es_conv}) may indicate an unrobust analysis method rather than genuine finite-size effects.

\section{Impact point sampling}
\label{sec:sampling}

\begin{figure}[h]
    \centering
    \includegraphics[trim={0.2in 0 0.2in 0},clip,scale=0.95]{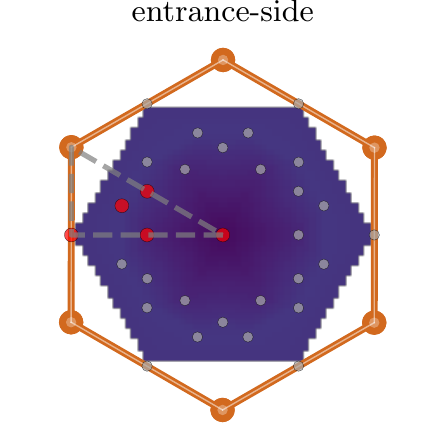}
    \includegraphics[trim={0.2in 0 0.2in 0},clip,scale=0.95]{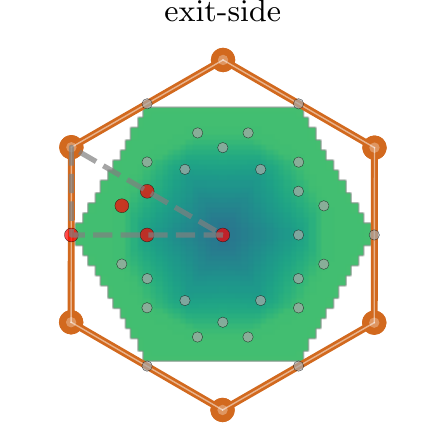}\\
    \includegraphics[trim={0.2in 0 0.2in 0},clip,scale=0.95]{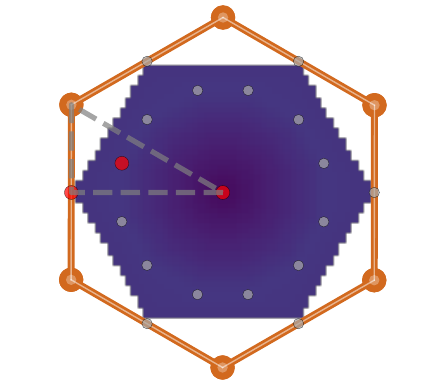}
    \includegraphics[trim={0.2in 0 0.2in 0},clip,scale=0.95]{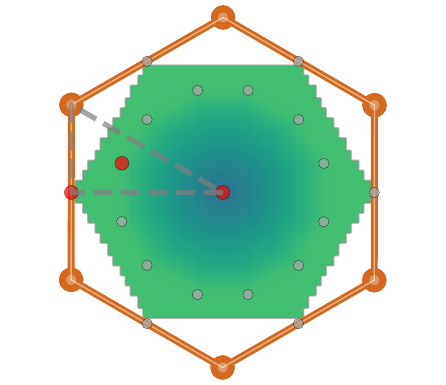}\\
    \includegraphics[scale=0.9]{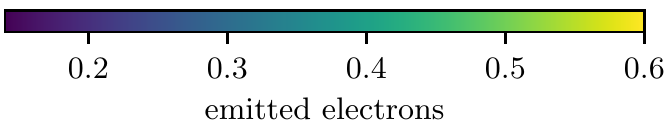}
    \caption{Simulated microscopy images for entrance-side (left) and exit-side (right) electron emission induced by 25\,keV protons impacting graphene (orange).
    The images were generated using 5 proton impact points (top) and only 3 representative proton impact points (bottom).
    Red points indicate explicitly simulated impact points within the gray, symmetry-irreducible triangle, and gray points indicate their symmetry equivalents.
    }
    \label{fig:traj_sampling}
\end{figure}

Accurate interpolation of simulated microscopy images requires sufficiently dense sampling of projectile impact points.
Fig.\ \ref{fig:traj_sampling} compares simulated microscopy images produced using different sets of impact points for the case of 25\,keV protons.
The images generated from only 3 impact points reproduce the essential features of the images generated from all 5 impact points simulated.
Therefore, only the 3 impact points illustrated in the bottom panels of Fig.\ \ref{fig:traj_sampling} were simulated for other beam parameters.

\section{Additional charge equilibration results}
\label{sec:extra_charge}

\begin{figure}
    \centering
    \includegraphics{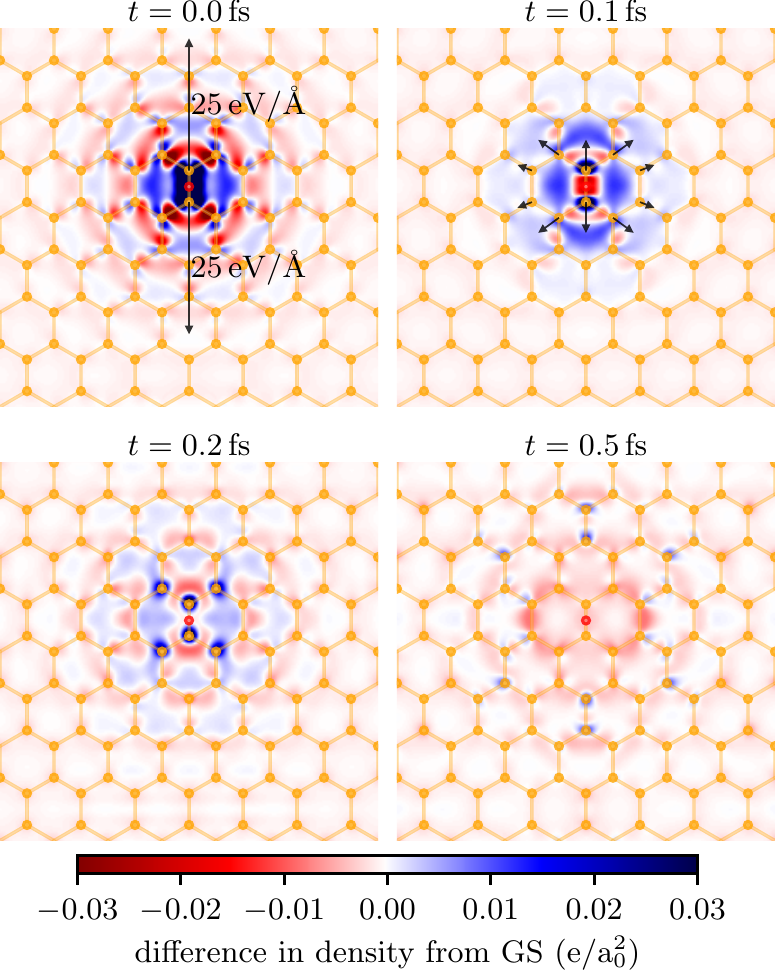}
    \caption{Snapshots of the charge distribution in graphene (orange) after a 50\,keV proton impacts at the midpoint of a C-C bond (red point).
    Red (blue) regions indicate lower (higher) electron density relative to the initial ground state, where the electron density has been integrated over a \SI{21}{\bohr}-thick slab centered on the graphene plane.
    In-plane atomic forces greater than \SI{2}{\electronvolt\per\angstrom} are indicated by black arrows.
    }
    \label{fig:h_charge_evolution}
\end{figure}

Here we present additional results related to charge dynamics within the irradiated graphene.
First, Fig.\ \ref{fig:h_charge_evolution} shows that after a proton impact, the charge in the material equilibrates on a sub-fs time scale, similar to the results reported for He$^{2+}$ ions in Fig.\ \ref{fig:charge_evolution} of the main text.
The charge distributions induced by both ions are qualitatively similar, but He$^{2+}$ ions cause higher magnitude deviations from the initial ground state.

\begin{figure}
    \centering
    \includegraphics{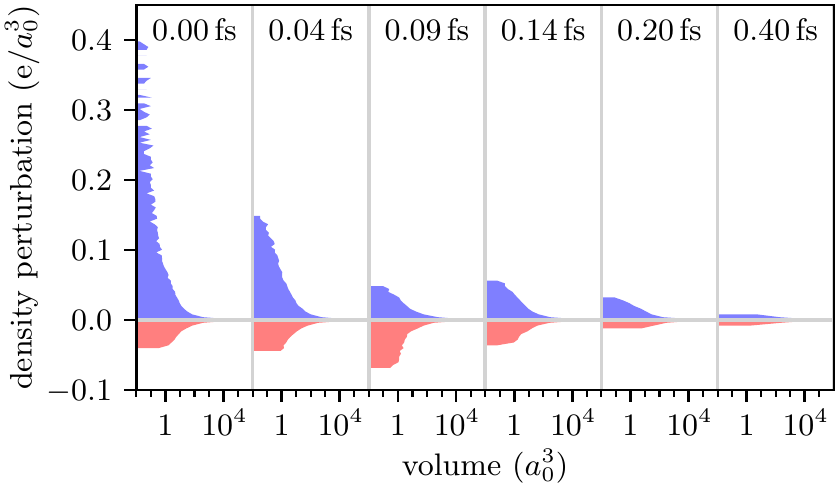}
    \caption{
    Snapshots of charge distribution relative to the initial ground state within a \SI{21}{\bohr}-thick slab centered on the graphene plane.
    Ion parameters are the same as in \mbox{Fig.\ \ref{fig:h_charge_evolution}}.
    Volume is given on a log scale, and labels along the top indicate time after impact.
    The distribution at each point in time is normalized to the total number of electrons within the region in question, which has a volume of $2.2\times 10^4$\,$a_0^3$ and initially contains 448 electrons.
    }
    \label{fig:charge_violin}
\end{figure}

A statistical analysis of the spatial charge distributions visualized in Figs.\ \ref{fig:charge_evolution} and \ref{fig:h_charge_evolution} can reveal additional information about the magnitude and extent of these dynamic charge perturbations.
Fig.\ \ref{fig:charge_violin} shows histograms of the time-dependent distribution of density perturbations relative to the ground state throughout the three-dimensional, \SI{21}{\bohr}-thick slab.
Both the maximum magnitude charge deviation and the distribution width again decrease on a sub-fs time scale, but do not behave monotonically over time as might be expected for simple diffusion of a localized charge distribution.

The integrals of of the positively and negatively charged portions of each distribution plotted in Fig.\ \ref{fig:charge_violin} estimate the numbers of excited electrons and holes, equivalent to Eq.\ \eqref{eq:charge_deviation} in the main text.
As an alternative approach, the number of excited holes can be derived from the time-dependent occupations of ground-state Kohn-Sham orbitals:
\begin{equation}
    N_{\mathrm{KS}}^{(\mathrm{h})}(t) = 2\sum_j \left[ 1 - \sum_\ell \left|\left\langle \phi_\ell(t) \left| \phi_j^{\mathrm{(GS)}}\right.\right\rangle\right|^2 \right],
    \label{eq:excited_KS}
\end{equation}
where $\phi_\ell(t)$ are time-dependent Kohn-Sham (KS) orbitals, $\phi_j^{\mathrm{(GS)}}=\phi_\ell(0)$ are initially occupied ground-state KS orbitals, and the factor of 2 accounts for spin degeneracy.
Since the electron density corresponding to the ground-state KS orbitals is almost completely confined to a \SI{21}{\bohr}-thick slab containing the graphene, $N_{\mathrm{KS}}^{(\mathrm{h})}(t)$ is analogous to the excited hole population computed from the electron density according to Eq.\ \eqref{eq:charge_deviation} and accompanying text and reported in Fig.\ \ref{fig:excited_es}a of the main text.

However, the total number of excited electrons, which is equal to $N_{\mathrm{KS}}^{(\mathrm{h})}(t)$ within this method, would include emitted and captured electrons that are not localized to the graphene.
Extracting the number of excited electrons remaining within the graphene using occupations of ground-state KS orbitals would require computing
\begin{equation}
    N_{\mathrm{KS}}^{(\mathrm{e})}(t) = 2\sum_{j',\ell} \left|\left\langle \phi_\ell(t) \left| \phi_{j'}^{\mathrm{(GS)}}\right.\right\rangle\right|^2 \int_{V_G} \left|\phi_{j'}^{\mathrm{(GS)}}\right|^2 \; dr^3,
    \label{eq:excited_e_KS}
\end{equation}
where $\phi_{j'}^{\mathrm{(GS)}}$ are initially unoccupied ground-state KS states and the integral incorporates the localization of each excited KS orbital within the analyzing slab $V_G$ containing the graphene.
We do not evaluate Eq.\ \eqref{eq:excited_e_KS} because it may require a large number of empty ground-state KS orbitals to converge.

\begin{figure}
    \centering
    \includegraphics{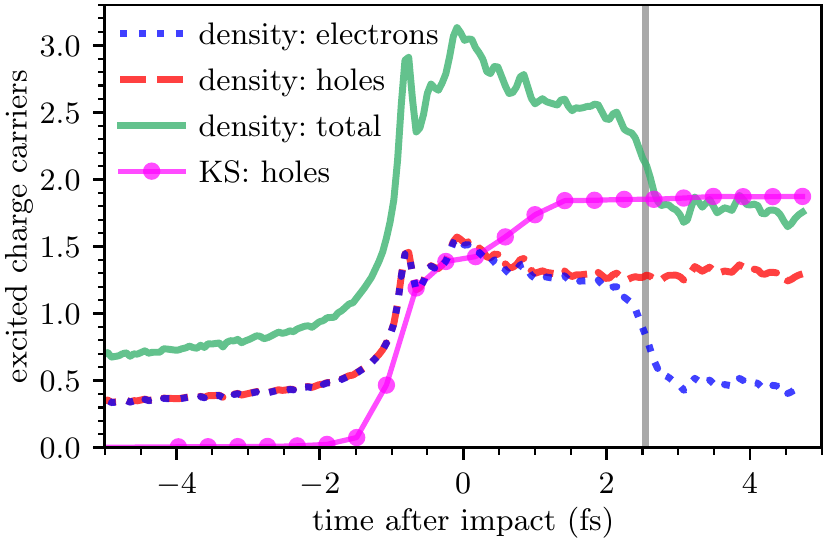}
    \caption{The number of excited electrons and holes after a proton with 0.1 atomic units of velocity impacts graphene along the centroid trajectory.
    Results computed using the electron density according to Eq.\ \eqref{eq:charge_deviation} (blue, red, and green curves) are compared to results from \mbox{Ref.\ \onlinecite{kononov:2021}} using the Kohn-Sham orbitals as given in Eq.\ \eqref{eq:excited_KS} (pink circles).
    The gray vertical line indicates when the projectile exits the \SI{21}{\bohr}-thick analyzing volume used in the electron density method.
    }
    \label{fig:excited_es_vs_KS}
\end{figure}

Nonetheless, we compare results from the density-based and orbital-based approaches in Fig.\ \ref{fig:excited_es_vs_KS}.
Although the number of excited holes predicted by the two methods exhibits similar qualitative behavior, we find quantitative differences depending on the regime.
First, more than \SI{1}{\femto\second} before impact, the density-based method predicts 0.3\,--\,0.5 excited holes while the orbital-based method gives nearly 0 excited holes.
We attribute this discrepancy to the inadequacy of the orbital-based method, which inherently posits a single-particle picture, in capturing the collective excitations that dominate at early times (e.g., sheet polarization, dynamical screening, and plasmon excitation).
Furthermore, Kohn-Sham orbitals are auxiliary constructs that generally do not have a rigorous physical significance, whereas their aggregate electron density does.

Single-particle excitations begin to occur when the proton enters the graphene's electron density within $\sim$\SI{1}{\femto\second} of impact, and the two methods agree quite well in this regime.
However, as the excited charge spreads and equilibrates after impact, the density-based method predicts about 30\% less holes than the orbital-based approach.
We interpret this discrepancy as a consequence of spatially superimposed valence band holes and excited electrons partially cancelling out within the total charge density, causing the density-based method to underestimate both the number of excited holes and the number of excited electrons.
After the projectile exits the analyzing volume along with captured and emitted electrons, leaving behind a positively charged graphene sheet with holes as the dominant excited charge carrier, the number of excited holes from the orbital-based method agrees with the total excited charge from the density-based method.
Given the deficiencies of both approaches, this agreement may simply be a coincidence.

\begin{figure}[h]
    \centering
    \includegraphics{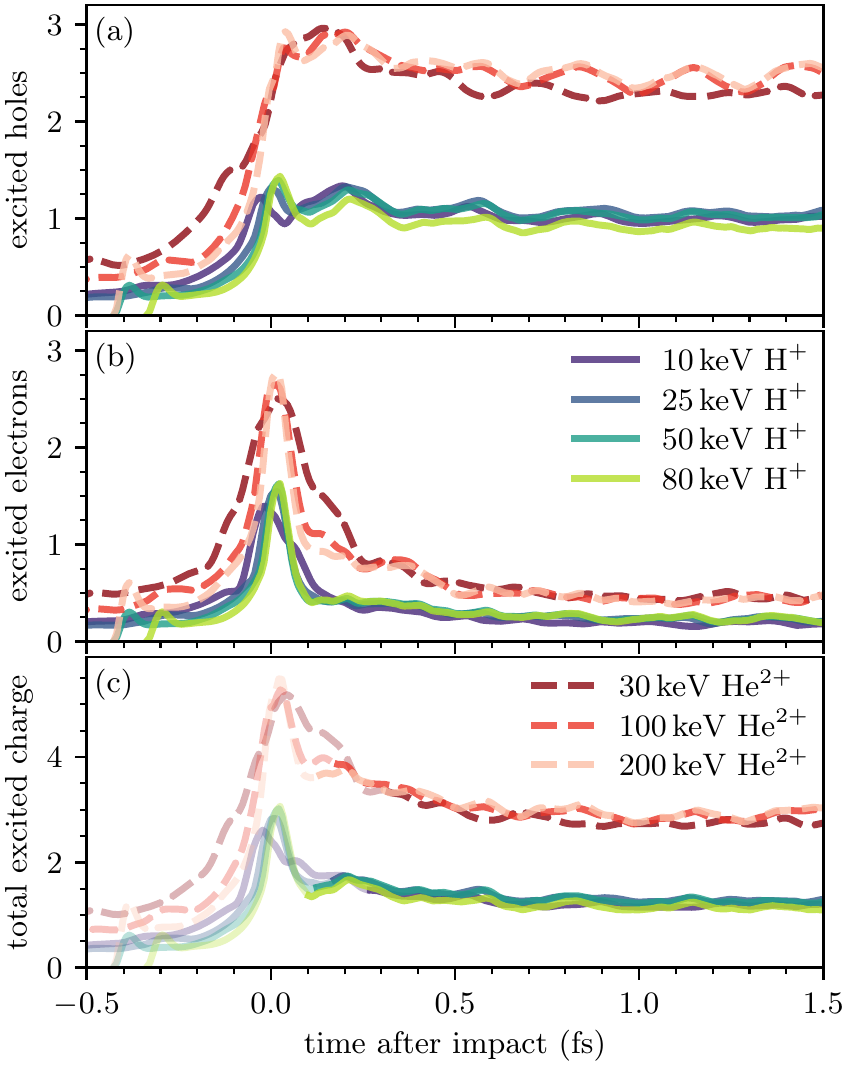}
    \caption{
    The number of excited holes (a), the number of excited electrons (b), and the total excited charge (c) in the irradiated graphene, as defined by Eq.\ \eqref{eq:charge_deviation} and accompanying text.
    Here, we use a \SI{4}{\bohr}-thick analyzing slab instead of the \SI{21}{\bohr}-thick slab considered in Fig.\ \ref{fig:excited_es} of the main text.
    In (c), light (dark) curves indicate data before (after) the cutoff time used in fitting to Eq.\ \eqref{eq:fit} for estimating the charge equilibration time.
    }
    \label{fig:excited_es_supp}
\end{figure}

Finally, we discuss the sensitivities of the characteristic charge equilibration times extracted according to Eq.\ \eqref{eq:fit} of the main text.
In addition to difficulties in converging this quantity with increasing graphene supercell size as discussed in Sec.\ \ref{sec:convergence}, results depend on the analyzing volume used to define excited charge in Eq.\ \eqref{eq:charge_deviation}.
The \SI{21}{\bohr} slab thickness used thus far for the analysis of charge dynamics in the graphene was chosen for consistency with the material-vacuum boundary used to determine emitted electron yields.
Of course, much of the graphene's electron density is concentrated within a thinner region.
Fig.\ \ref{fig:excited_es_supp} shows the number of electrons excited within a \SI{4}{\bohr}-thick slab, analogous to Fig.\ \ref{fig:excited_es} in the main text.
The amount of excited charge, particularly the number of excited electrons, begins to decay sooner after impact in Fig.\ \ref{fig:excited_es_supp} than in Fig.\ \ref{fig:excited_es} because the captured and emitted electrons emerge from the thinner slab earlier.
Otherwise, the qualitative behavior remains unchanged.
However, a \SI{4}{\bohr}-thick analyzing slab produces equilibration times up to 43\% or \SI{0.1}{\femto\second} shorter than a \SI{21}{\bohr}-thick analyzing slab, a difference much greater than typical fit uncertainties around \SI{0.01}{\femto\second}.
Altering the cutoff time for the data included in the fit to Eq.\ \eqref{eq:fit} from the time when the projectile has traveled \SI{5}{\bohr} away from the analyzing slab to the time when the projectile has traveled \SI{3}{\bohr} away from the analyzing slab also changes the extracted equilibration times by up to 24\% or \SI{0.03}{\femto\second}.
Despite these large relative uncertainties, our predicted equilibration times should have the correct order of magnitude and qualitative behavior wherein equilibration times increase with ion charge.

\section{Additional analysis of atomic forces}
\label{sec:extra_forces}

\begin{figure}[h]
    \centering
    \includegraphics{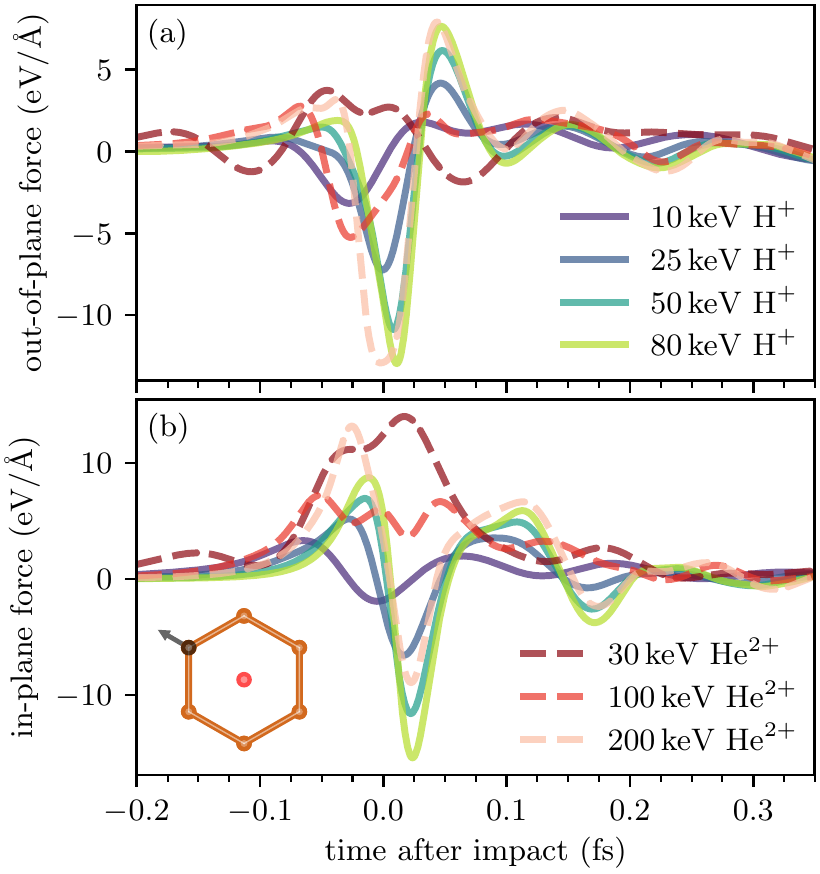}
    \caption{Instantaneous out-of-plane (a) and in-plane (b) force experienced by a nearest C atom after an ion impacts as shown in the inset.
    Positive out-of-plane forces point in the direction of projectile motion, and positive in-plane forces point away from the impact point.
    }
    \label{fig:forcesA}
\end{figure}

\begin{figure}
    \centering
    \includegraphics{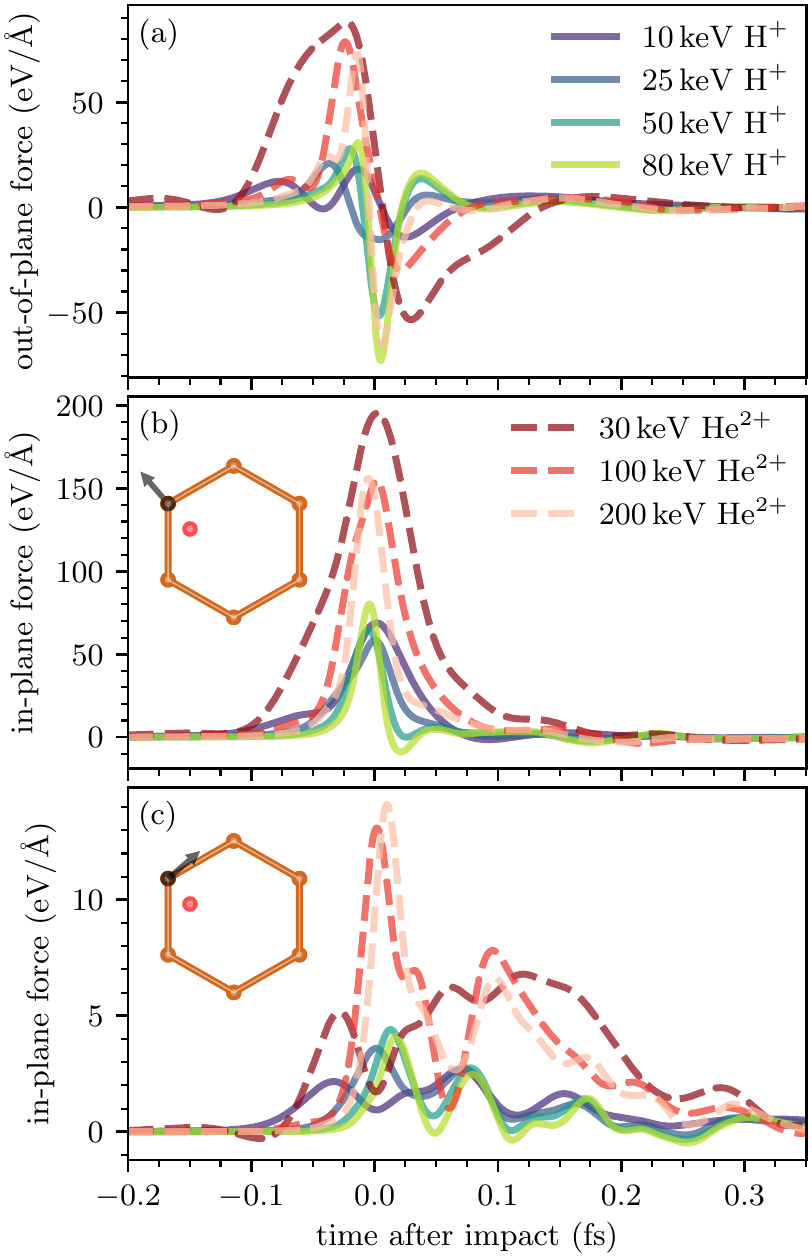}
    \caption{Instantaneous out-of-plane (a) and in-plane (b, c) forces experienced by a nearest C atom after an ion impacts as shown in the inset.
    Positive out-of-plane forces point in the direction of projectile motion, and positive in-plane forces point as indicated by black arrows in the insets.
    }
    \label{fig:forcesO}
\end{figure}

The discussion of Sec.\ \ref{sec:damage} in the main text focused on ions impacting C\,--\,C bonds.
In Figs.\ \ref{fig:forcesA} and \ref{fig:forcesO}, we report results analogous to Fig.\ \ref{fig:forcesF} of the main text for other impact points.
While the behavior in Fig.\ \ref{fig:forcesO} is very similar to the discussion of Fig.\ \ref{fig:forcesF}, ions impacting at the center of a carbon hexagon induce considerably smaller forces on nearby C atoms (see Fig.\ \ref{fig:forcesA}).
Although the signs of the forces in Fig.\ \ref{fig:forcesA} deviate from the trends observed for other ion impact points, the net impulses do not: very small impulses corresponding to at most \SI{2}{\milli\electronvolt} of kinetic energy transfer work to displace the nearest C atoms away from the impact point.
We note that the lack of symmetry in the case of the impact point considered in Fig.\ \ref{fig:forcesO} also induces a transverse in-plane force, but its magnitude is about 10 times smaller than the force pointing away from the impact point (see \mbox{Fig.\ \ref{fig:forcesO}b and c}).

\end{document}